\documentclass[twocolumn,pre]{revtex4}

\usepackage{graphicx}
\usepackage{dcolumn}
\usepackage{bm}

\usepackage{graphics}
\begin{document}

\title{Effective charge and free energy of DNA inside an ion channel}

\author{Jingshan Zhang}
\author{B. I. Shklovskii}
\affiliation{Theoretical Physics Institute, University of
Minnesota, Minneapolis, Minnesota 55455}

\date{\today}

\begin{abstract}

Translocation of a single stranded DNA (ssDNA) through an
$\alpha$-hemolysin channel in a lipid membrane driven by applied
transmembrane voltage $V$ was extensively studied recently. While
the bare charge of the ssDNA piece inside the channel is
approximately $12$ (in units of electron charge) measurements of
different effective charges resulted in values between one and
two. We explain these challenging observations by a large
self-energy of a charge in the narrow water filled gap between ssDNA
and channel walls, related to large difference between dielectric
constants of water and lipid, and calculate effective charges of
ssDNA. We start from the most fundamental stall charge $q_{s}$,
which determines the force $F_{s}= q_s V/L$ stalling DNA against
the voltage $V$ ($L$ is the length of the channel). We show that
the stall charge $q_{s}$ is proportional to the ion current
blocked by DNA, which is small due to the self-energy barrier.
Large voltage $V$ reduces the capture barrier which DNA molecule
should overcome in order to enter the channel by $|q_{c}|V$, where
$q_{c}$ is the effective capture charge. We expressed it through
the stall charge $q_{s}$. We also relate the stall charge $q_{s}$
to two other effective charges measured for ssDNA with a hairpin
in the back end: the charge $q_u$ responsible for reduction of the
barrier for unzipping of the hairpin and the charge $q_e$
responsible for DNA escape in the direction of hairpin against the
voltage. At small $V$ we explain reduction of the capture barrier
with the salt concentration.

\end{abstract}

\maketitle

\section{Introduction}

A DNA molecule in a water solution carries negative charges. With
the help of applied voltage, it can translocate through a wide
enough ion channel located in a lipid
membrane~\cite{Henrickson,Meller2001,Meller2002,Meller2003,Sauer,Mathe,Ambjörnsson,Nakane,Bonthuis}
or through a solid state nanopore in a semiconductor film
\cite{Li,Lemay,Min}. An intensively studied example is the
translocation of a single stranded DNA (ssDNA) molecule through an
$\alpha$-hemolysin ($\alpha$-HL) channel~
\cite{Henrickson,Meller2001,Meller2002,Meller2003,Sauer,Mathe,Ambjörnsson,Nakane,Bonthuis}.
With the average internal diameter $\sim 1.7\,$nm the channel is
wide enough for ssDNA molecules, but is too narrow for a double
helix. To be specific below we always talk about the experimental
data for this system. Our theory is also applicable to a double
helix DNA translocating through a narrow nanopore
\cite{Li,Lemay,Min}, but there is less quantitative data for this
case.

\begin{figure}[ht]
\begin{center}
\vspace{-3.5cm}
\includegraphics[width=8.5cm, keepaspectratio]{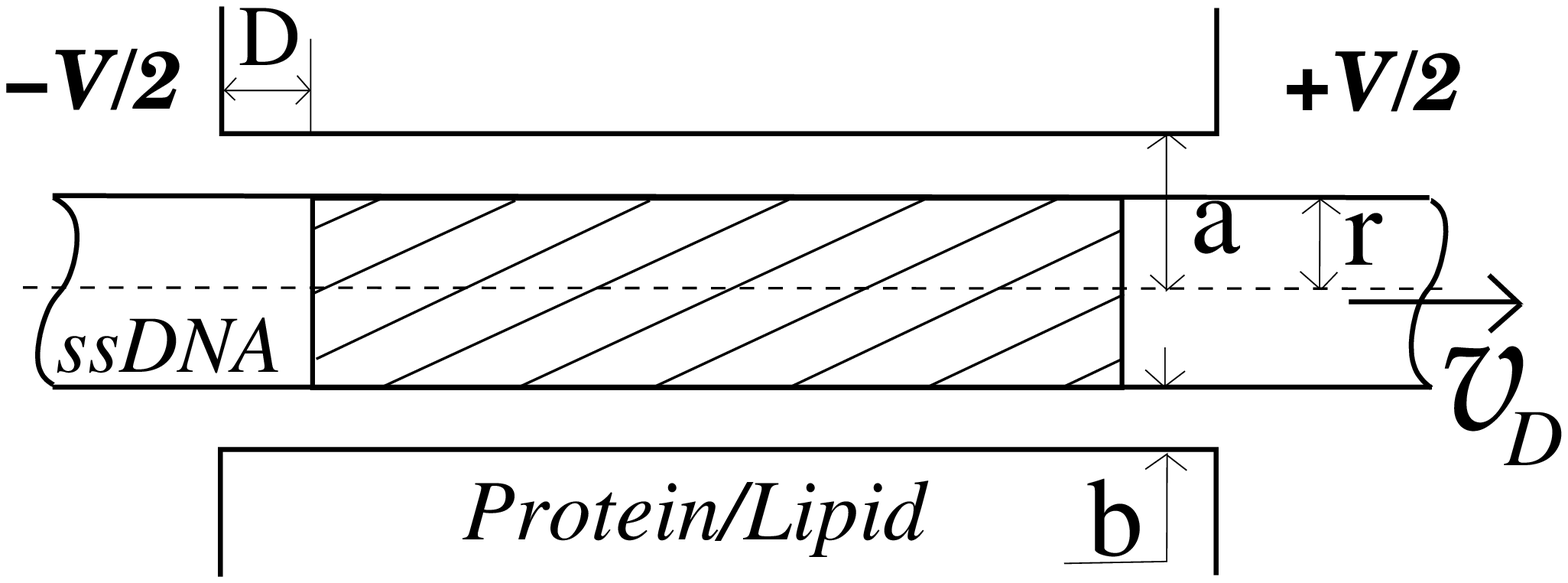}
\vspace{-3.5cm}
\end{center}
\caption{The side view of the membrane and the channel with captured
DNA. All DNA phosphates in the shaded part are neutralized by K$^+$
ions. The contact layer on each end has the length $D$. The arrow
shows the direction of DNA motion.} \label{figdna}
\end{figure}

In order to study the translocation experimentally, the electric
current through the channel is observed under voltage $V$, applied
between two vessels of salty water on both sides of the membrane.
Due to large conductivity of the bulk solution practically all the
voltage drops on the membrane. When a ssDNA is added to the
negative voltage side it is dragged into the channel by the
voltage (Fig.~\ref{figdna}). When the ssDNA molecule is in the
channel as shown in Fig.~\ref{figdna} the ion current is blocked,
and the blocked current $I_b$ is much smaller than the open pore
current $I_0$ without DNA in it,
\begin{equation}
I_b \simeq 0.1\, I_0. \label{blockage}
\end{equation}
Translocation events in a single channel can be studied by
monitoring the current. It was argued recently that the steric
mechanism of strong current blockage is amplified by the increase
electrostatic self-energy of an ion in water passage narrowed by
DNA~\cite{Bonthuis} because DNA and lipids have much smaller
dielectric constants than water and, therefore, the electric field
lines of an ion in the space between DNA and the lipid are
squeezed in the channel increasing the ion self-energy. This idea
was borrowed from the physics of narrow ion channels without
DNA~\cite {Parsegian}.

Besides the blocked current, one can measure the time between two
successive translocation events, $\tau$, or the capture rate $R_c
= 1/\tau$ of DNA molecules into the channel. It is natural to
compare the observed value of $R_c$ with the diffusion limited
rate $R_D$ of ssDNA capture. This comparison shows that $R_c \ll
R_D$. For instance~\cite{Meller2003}, the typical $R_c$ is in the
range of $0.01 - 10\,s^{-1}$ at applied voltage $50 -
200\,$mV and ssDNA concentration $0.9\,\mu$M, while $R_D \sim 100
s^{-1}$. The ratio $R_c / R_D$ may be as small as $10^{-6}$ if one
extrapolates the experimental data to $V=0$. So there must be a
large barrier $\sim 14 k_BT$ for ssDNA capture. We return to the nature
of this
barrier in the end of Introduction, but first we concentrate on
challenging question of the voltage effect on this barrier.

The capture rate at zero voltage $R_{c}(0)$ is so small that all
experiments are actually done with a large applied voltage $V = 50
- 200$ mV . The voltage pulls DNA into the channel and reduces the
barrier for DNA capture. It was
found~\cite{Henrickson,Meller2002,Meller2003} that the capture
rate is
\begin{equation}
R_{c}(V) = R_{c}(0) \exp (- q_{c}V/k_B T ), \label{caprate}
\end{equation}
where $ q_c = - 1.9e$ is an effective ``capture" charge, and $e$
is the proton charge. Apparently for ssDNA in $\alpha$-HL channel
$|q_c|$ is much smaller than the absolute value of the total DNA
charge in the channel $-N_{0}e \simeq -12e$ and this is why large
voltage $V \gg k_BT/e$ is necessary~\cite{Meller2001} in order to
make the capture rate observable.

Why is the capture charge of DNA so small and what does it depend
upon? How is the capture charge related to the stall charge
defining the stall force $F_s$, which one should apply to DNA
occupying the whole length of the channel (Fig.~\ref{figdna}) to
stall it against the voltage $V$? (This, for example, can be done
with the help of an laser tweezers~\cite{Smith,Lemay}). In other
words, $F_{s} = - F_{p}$, where $F_p$ is the force, with which the
voltage $V$ pulls the stalled DNA. We write $F_s$ as
\begin{equation}
F_{s} = - q_{s}E, \label{fpqp}
\end{equation}
where the electric field $E = -V/L$ and $q_s$ is the stall
effective charge. The charge $q_{s}$ seem to be the simplest and
the most fundamental effective charge one can introduce for DNA.
Is it different from $ q_{c}$? If yes, which one is larger? How
are these two charges related to unzipping and escape charges
which describe ssDNA with a hairpin at the end (see definitions
below)?

Inspired by all these challenges in this paper we use for DNA the
simplest model of a rigid cylinder charged by the point like
surface charges (phosphates located on the spiralling backbone)
and moving coaxially through a cylindrical tunnel filled by salty
water (Fig.~\ref{figdna}). The model of rigid cylinder should be
good for a double helix DNA in a narrow cylindrical semiconductor
pore. For ssDNA such model gets some support from the known
tendency of ssDNA to stuck its bases~\cite{Saenger1984} in a bulk
solution. It is natural to expect that this tendency is enhanced
inside the channel. One may say that the case of ssDNA in
$\alpha$-HL channel pushes our model too close to the molecular
limit. Nevertheless, we will show that our results for effective
charges are in a reasonable agreement with experiment. We believe
that this happens because these results are practically model
independent.

The plan of this paper is as follows. In
Sec.~\ref{sec_Electrostatics} we discuss our model and the
electrostatics of the channel. We concentrate on the role of the
enhanced self-energy of a salt ion in the narrow water-filled space
between DNA and internal walls. First, we argue that DNA in the
channel is almost perfectly neutralized and, second, salt cations
are bound to DNA charges. Then we introduce narrow charged contact
layers near the end of DNA and qualitatively explain the
electrostatic mechanism of the current blockage.

In Sec.~\ref{sec_qs} we calculate the stall force $F_s$ and the
stall effective charge $q_s$. Our main result is that $q_s$ is
proportional to the ratio of currents in blocked and opened
channels
\begin{equation}
q_{s}\simeq -eN_{0}\frac{I_b}{I_0}. \label{main}
\end{equation}
Using Eq.~(\ref{blockage}) we get that for ssDNA in $\alpha$-HL
channel stall charge $q_{s}\sim -1e$. Intuitively this is clear
because if the blocked current were exactly zero this would mean
that counter ions are stuck on DNA and, therefore, compensate the
DNA charge, so that the net pulling charge would vanish.
\begin{figure}[ht]
\begin{center}
\vspace{-3.5cm}
\includegraphics[width=8.5cm, keepaspectratio]{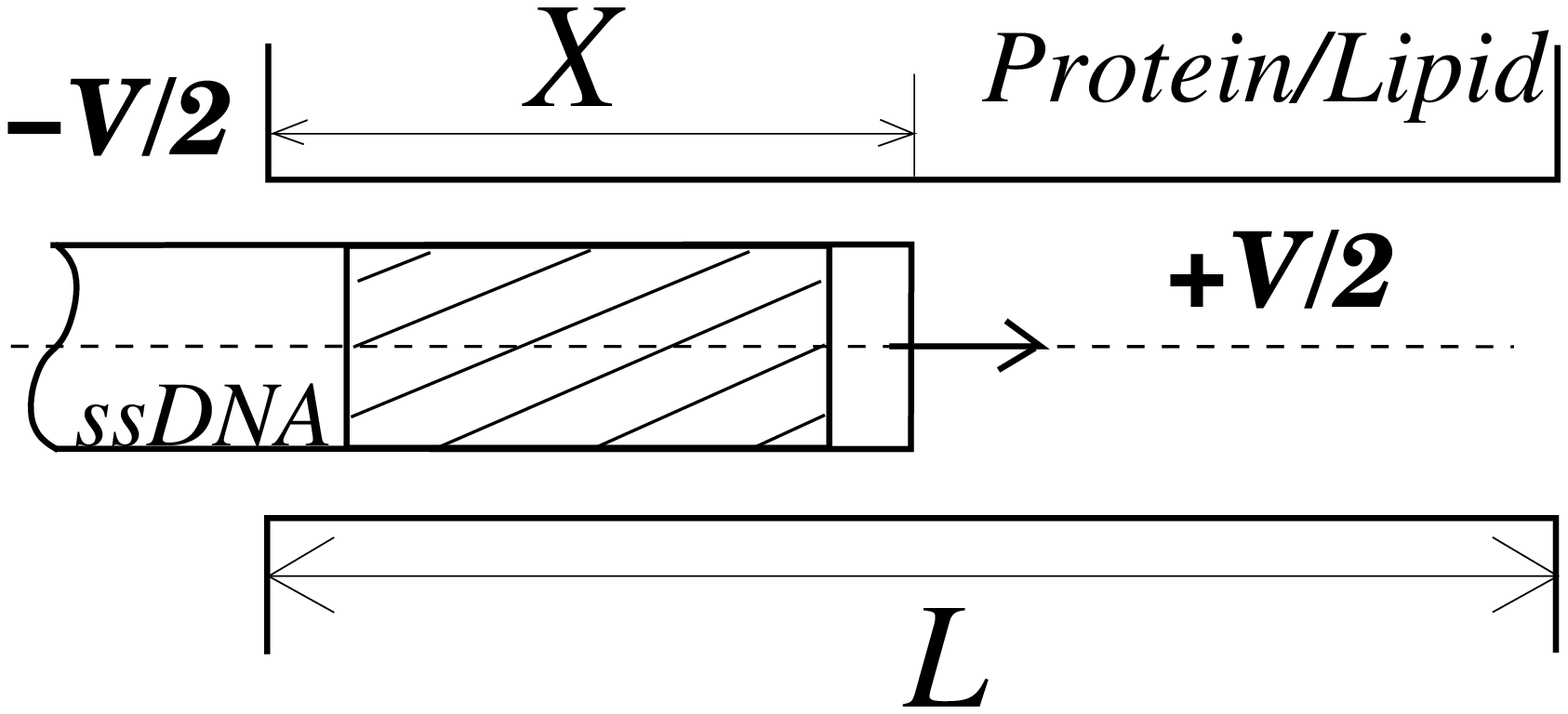}
\vspace{-3.9cm}
\end{center}
\caption{The side view of a ssDNA molecule entering the channel.}
\label{figdnapart}
\end{figure}

In Secs.~\ref{sec_qc} and ~\ref{sec_qu} we show that the stall
charge is the fundamental charge so that all other charges can be
expressed in terms of it. In Sec.~\ref{sec_qc} in order to find
the capture charge $q_c$ we calculate of the pulling force
$F_{p}(X)$ and the stalling force $F_{s}(X)= - F_p(X)$ for the
partial penetration of ssDNA into the channel to the depth $X < L$
(see Fig.~\ref{figdnapart}). Then the correction to the capture
barrier is calculated as a work which the electric field $E=-V/X$
does slowly pulling DNA into the whole channel. In order to obtain
the voltage correction $-q_{c}V$ to the minimum work necessary to
overcome the capture barrier we integrate the pulling force over
$X$. The resulting capture charge $q_c$ is larger than the stall
charge $q_{s}$. The reason is that the self-energy barrier becomes
smaller for a shorter channel.

\begin{figure}[ht]
\begin{center}
\vspace{-3.5cm}
\includegraphics[width=8.5cm, keepaspectratio]{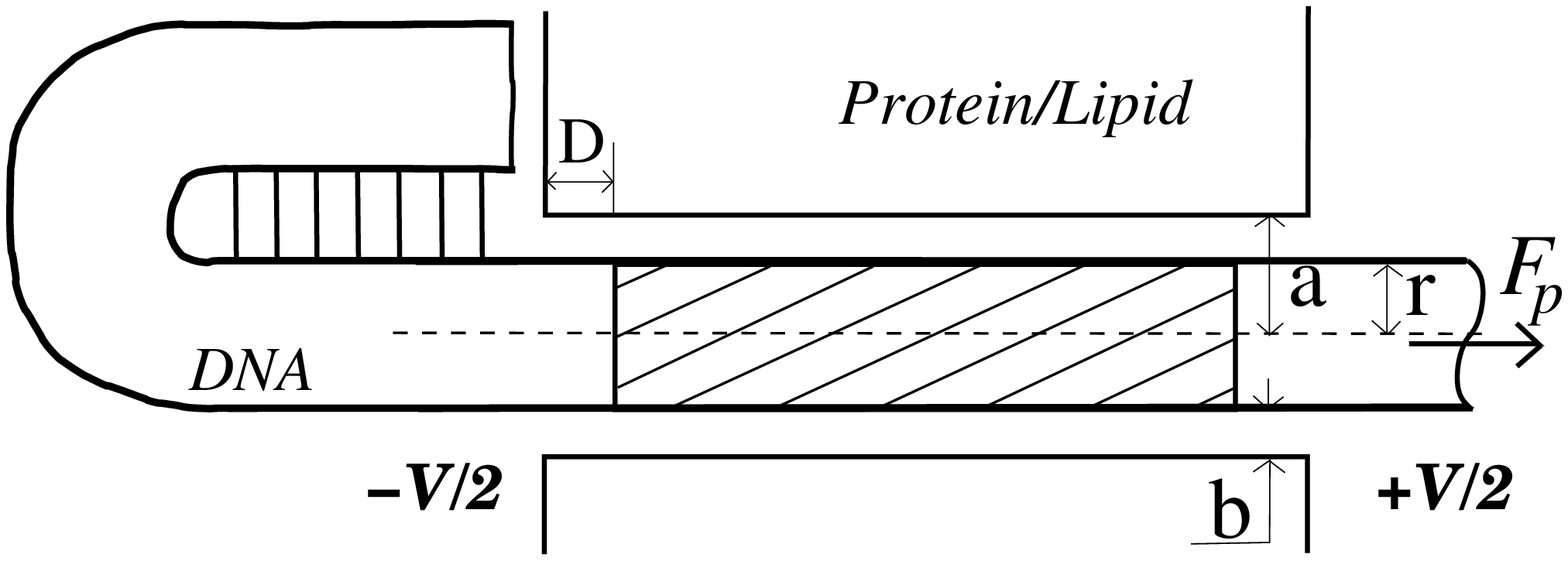}
\vspace{-4.5cm}
\end{center}
\caption{Unzipping of the DNA hairpin with the help of the voltage
induced pulling force $F_p$ results in DNA translocation through
the channel to the right. Alternatively DNA can escape against
electric field to the left. The hairpin is shown schematically
with the bound base pairs presented by short straight lines. There
are $M$ base pairs in the hairpin, $N_0$ bases in the channel and
$K$ bases in the tail to the right of the channel.}
\label{fighairpin}
\end{figure}

In Sec.~\ref{sec_qu} we discuss effective charges, which have to
do with release rate of ssDNA, when it is trapped in the channel
due to a hairpin in the back end (see Fig.~\ref{fighairpin}).
There are two ways for such a DNA molecule to leave the channel:
DNA can get unzipped by pulling electric field and leave to the
right or DNA can escape against the pulling force of the electric
field to the left. The former route dominates at large voltages,
while the latter one dominates at smaller ones.

It was found~\cite{Sauer,Mathe} that the unzipping rate
exponentially grows with the voltage as
\begin{equation}
R_{u}(V)=R_{u}(0)\exp(|q_{u}|V/k_B T). \label{escaperate}
\end{equation}
We show that the ``unzipping" effective charge $q_{u}=
q_{s}(M/N_0)$, where $M$ is the number of base pairs in the
hairpin. For used in \cite{Mathe} $M \simeq 10$ and $N_0 \simeq
12$, Eq. (\ref{escaperate}) gives $q_u \approx q_s \approx -1e$ in
agreement with \cite{Sauer,Mathe}.

The rate of alternative escape against the voltage should
exponentially decrease with $V$
\begin{equation}
R_{e}(V)=R_{e}(0)\exp(- q_{e}V/k_B T). \label{relrate}
\end{equation}
Here $q_{e}$ is the fourth effective charge, which we call the
``escape" charge. We show in Sec.~\ref{sec_qu} that $q_{e}=
|q_{s}|(K/N_0)$, where $K$ is number of bases in the ssDNA tail on
the right side of the membrane when escape begins.

In Sec.~\ref{sec_barrier} we are concerned with the nature of the
capture rate barrier. Reduction of the conformation entropy due to
confinement of the DNA piece in the channel was suggested as a
natural explanation for the capture rate barrier
\cite{Ambjörnsson}. Indeed, ssDNA molecules in the bulk solution
are rather flexible, with the persistence length about $p \simeq
1.4\,$nm \cite{Murphy} at $1$M KCl. The channel length $L \simeq
5\,$nm, so it holds $N_p=L/p\simeq 3.5$ persistence lengths of
ssDNA during translocation, and their undulations are restricted
by the channel. The large entropic barrier due to this effect is
$N_{p}k_{B}T\Delta s $, where $\Delta s$ is the loss of entropy
for one persistence length in the channel. Using $\Delta s\sim 2$
we get $\sim 7\,k_{B}T$ for this barrier.

An additional entropy loss comes from free tails of DNA outside
the channel \cite{Sung}. When one end of the DNA chain is anchored
onto the wall, the Gaussian chain has the free energy
${1\over2}\ln M_p$, where $M_p$ is the length of the tail in
persistence lengths. The ssDNA molecules used in experiments
\cite{Henrickson,Meller2003} are relatively short ($\le 40$
bases), and two tails can give only a barrier $\sim 1\!-\!2
\,k_{B}T$. All the losses of conformation entropy together can
explain a substantial part of the estimated barrier $\sim
14\,k_{B}T$ extrapolated to zero voltage at salt concentration
$1$M KCl.

However, they cannot explain the observed dependence of the
capture rate on the salt concentration $c$. Indeed, the
persistence length of ssDNA decreases when $c$ increases
\cite{Murphy}, making the conformation barrier larger, while in
the experiment the capture rate grows with $c$~\cite{Bonthuis}. So
there must be another kind of barrier with the opposite $c$
dependence.

In Sec.~\ref{sec_barrier} we suggest a mechanism for such a
barrier. We argue that when a DNA molecule enters the channel, the
screening cloud is squeezed in the narrow water-filled space
surrounding the DNA. Due to this compression the total free energy
of DNA and ions is higher for DNA in the channel than for DNA in
the bulk. This barrier decreases with $c$ because entropy of
screening atmosphere in the bulk solution decreases. Using even
more simplified model of DNA as uniformly charge cylinder and the
Poisson-Boltzmann approximation we show that this barrier is in
qualitative agreement with the observed dependence $R_c$ on the
salt concentration $c$.

In Sec.~\ref{sec_conclusion} we conclude with the summary of our
results.

\section{Neutralization of DNA in the channel and the contact potential}
\label{sec_Electrostatics}

We assume the ssDNA molecule is a rigid cylinder coaxial with the
channel. The inner radius of the $\alpha$-HL channel is
$a\!\simeq\! 0.85\,$nm, and the radius of the ssDNA molecule is
$r\!\simeq\! 0.5\,$nm (Fig.~\ref{figdna}). Salt ions are located
in the water-filled space between them, with thickness
$b\!\simeq\! 0.35\,$nm. The length of the channel is $L\!\simeq\!
5\,$nm. Such a model is even more appropriate for double helix DNA
in a wider (say $4\,$nm in diameter) solid state nanopore
\cite{Li,Min}.

\begin{figure}[ht]
\begin{center}
\vspace{0cm}
\includegraphics[width=8.5cm, keepaspectratio]{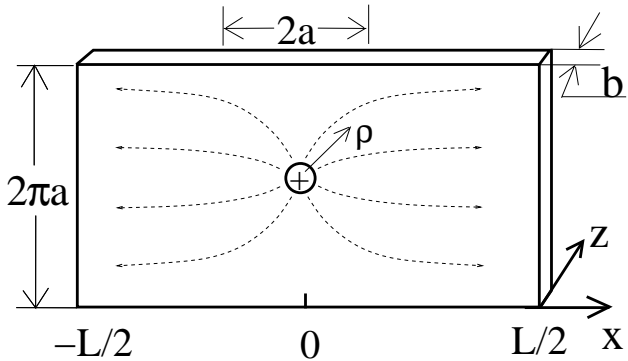}
\vspace{-4.9cm}
\end{center}
\caption{An unfolded view of the water-filled space containing an
extra K$^+$ ion. One side of the water-filled space between ssDNA
and the channel walls is cut by a radial half plane starting from
the channel axis and the cut is unfolded to make the water filled
space flat. Dashed lines represent the electric field lines of the
charge. At $\rho < a$ this electric field spreads in all
directions and becomes uniform far from the charge.}
\label{figfield}
\end{figure}
The dielectric constant of the channel and the ssDNA molecule
($\kappa'\!\sim\! 2$) is much smaller than that of water
($\kappa\!\simeq\! 80$). So if ssDNA is neutralized by cations and
there is an extra charge $e$ at the point $x$ located in the thin
water-filled space between the channel internal wall, the electric
field lines starting from this charge are squeezed in the thin
layer (Fig.~\ref{figfield}). This results in a high self-energy
$U(x)$ of the charge~\cite{Parsegian,Zhang}.

\begin{figure}[ht]
\begin{center}
\vspace{-2cm}
\includegraphics[width=7.5cm, keepaspectratio]{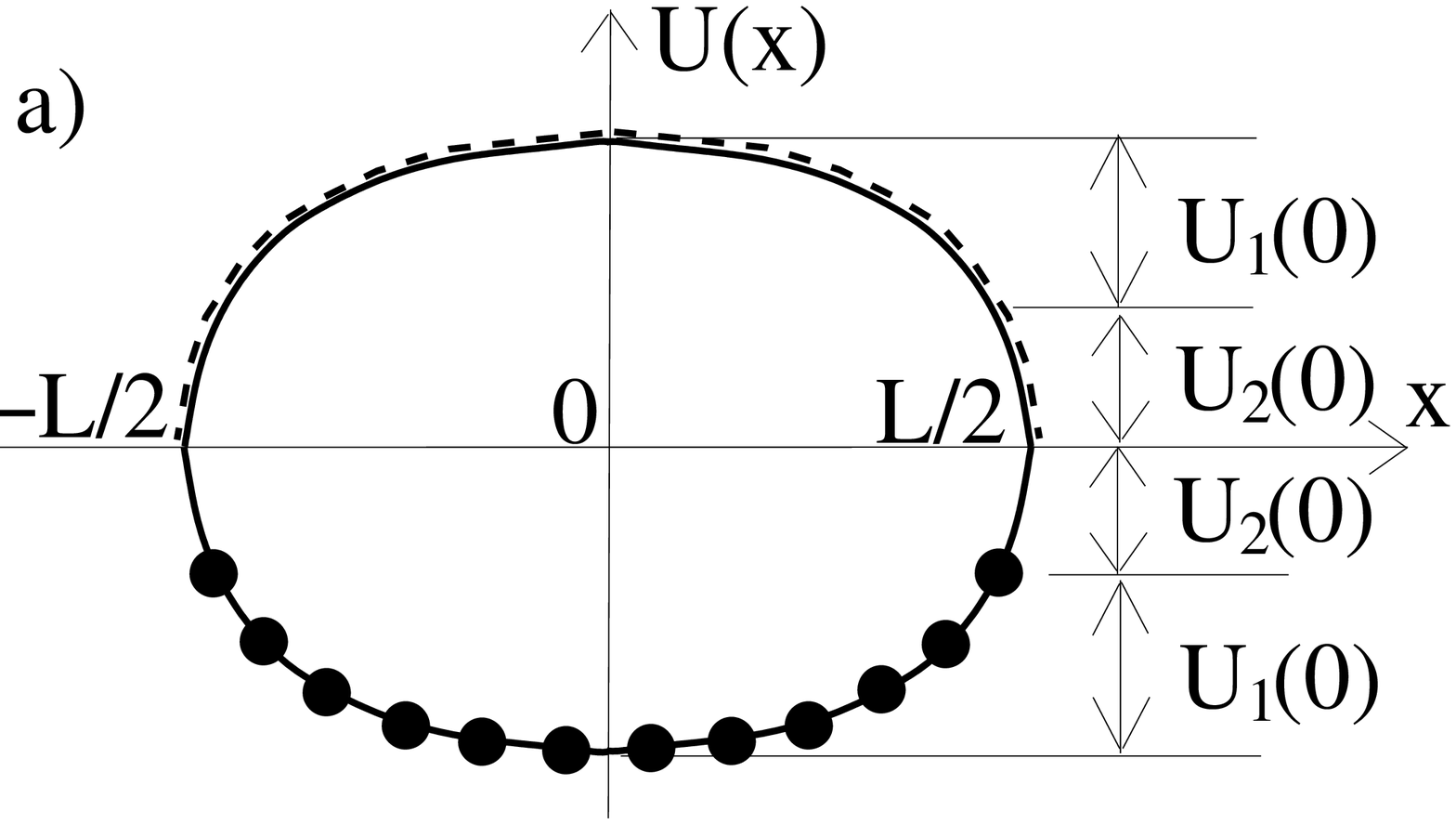}
\end{center}
\begin{center}
\vspace{-4cm}
\includegraphics[width=7.5cm, keepaspectratio]{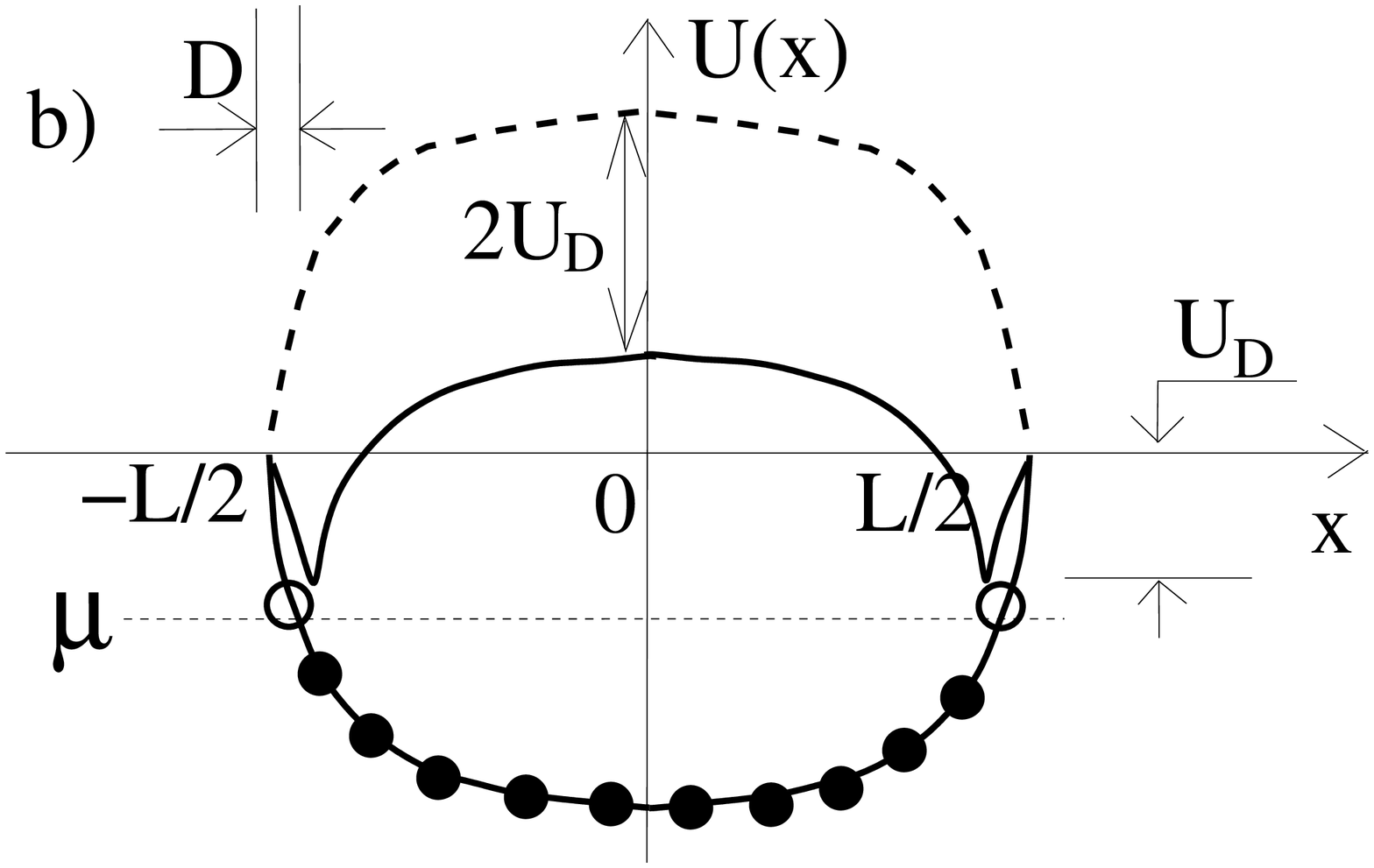}
\vspace{-2cm}
\end{center}
\caption{Energy band diagram for K$^+$ ions (solid lines) and
Cl$^-$ ions (dashed lines). The lower band represents the energy
of the cations bound to DNA phosphates (the self-energy necessary
to create a vacancy with sign minus). The empty upper bands show
the self-energy of the extra salt cation (solid line) and anions
(dashed line) entering the channel. a) In the absence of the
contact layers ($c > c_D$). b) with contact layers of the width
$D$ creating contact electrostatic potential $-U_D$. Vacant
phosphates are shown by empty circles. The chemical potential
$\mu$ of K$^+$ ions in the system is shown by the thin dotted
line.} \label{figband}
\end{figure}

In order to calculate $U(x)$ we write $U(x) = e\phi(x)/2$, where
$\phi(x)$ is the electrostatic potential created by the extra
charge at position $x$. If $\rho$ is the distance from the charge
$e$, electric field is two-dimensional at $\rho < a$ (see
Fig.~\ref{figfield}), and becomes uniform at larger distance $\rho
 > a$. Our numerical calculation in the limit of infinite ratio
$\kappa/\kappa'$, when all electric lines stay in the channel and
at $a/b\gg 1$ can be well approximated by the following
expression:
\begin{equation}
U(x) = U_{1}(x)\!+\!U_{2}(x) =\frac{e^{2}}{\kappa
b}\left[\frac{L}{4a}\left(\!1\!-\!{4x^2 \over L^2}\!\right) \!+\!
\ln\frac{a}{b}\right]. \label{barrier}
\end{equation}
The origin of the two terms in Eq.~(\ref{barrier}) is illustrated
in Fig.~\ref{figfield} for $x=0$. At $b < \rho < a$ the electric
field of the central charge gradually spreads over all azimuthal
angles in the the whole water-filled space decaying as
$E=2e/\kappa\rho b$. This leads to the two-dimensional potential
$\phi(x) = (2e/\kappa b)\ln(a/b)$ and produces the constant term
$U_{2}(x)$ in Eq.~(\ref{barrier}). ( Eq.~(\ref{barrier}) works
when the charge is not too close to the channel ends, $L/2-|x|
 >a$. However, when the charge is at channel ends
Eq.~(\ref{barrier}) is no longer valid. Instead the electric field
lines are attracted to the bulk solution, and $U_{2}(x)$
vanishes.) On the other hand, $U_{1}(0)$ is created by the
one-dimensional uniform electric field at distances $\rho > a$,
which according to Gauss' theorem is $E_0 = e/\kappa ab$. For
$|x|>0$ the electric field at the closer end is stronger than that
of the other end, therefore $U_1(x)$ decreases parabolically with
$|x|$, and vanishes at the channel ends~\cite{Kamenev}. For $L= 5$
nm, $a=0.85$ nm and $b= 0.35$~nm Eq.~(\ref{barrier}) gives
$U_{1}(0)= 2.9 k_BT$ and $U_{2}(0) = 1.7 k_BT$ where $T$ is the
room temperature. The total barrier $U(x)$ of an extra K$^+$ or
Cl$^-$ ion is shown on Fig.~\ref{figband} by the upper thick
line~\cite{foot}.

Now recall that there are K$^+$ ions bound to ssDNA phosphates in
the channel. Each of them can be removed to the bulk solution
creating a vacancy. The energy penalty for this process is close
to the penalty for placing an extra ion in the same place. Thus,
energies of bound K$^+$ ions are $- U_{1}(x) -U_2$ and can be
shown by the lower full curve of Fig.~\ref{figband} as a
reflection of the upper one with respect of the $x$ - axis.
Vacancies have to overcome the barrier $U(0)$ to cross the
channel. Using an analogy with semiconductors, we can say that
extra K$^+$ ions play the role of electrons in the conduction
band, while vacancies play the role of holes in the valence band.
The peculiar result of electric field confinement in the
water-filled space is that the energy gap $2(U_{1}(x) +U_2)$ has
the maximum at $x=0$ (Fig.~\ref{figband}). In the above discussion
we ignored entropy effects~\cite{Zhang,Kamenev}, which in
principle can reduce self-energy barriers~\cite{foot1}.

The most important for us conclusion from above discussion is that
the large self-energy of extra charges deep inside the channel
leads to very accurate neutralization of DNA by salt cations. Such
nearly perfect neutralization was observed in computer simulations
~\cite{Rabin} of the channel.

When salt concentration in the bulk solution $c$ is smaller than the
characteristic concentration of K cations $c_D$ in DNA occupied
channel, some K cations close to the channel ends can escape to the
bulk in order to enjoy larger entropy in the solution. As a result
there are negative phosphate charges in the layer of width $D$ at
each end, and the screening (positive) charge in the adjacent layers
of the bulk solution. These double layers (capacitors) of the width
$D$ (see Fig.~\ref{figdna}) produce the contact potential $-U_D$,
where
\begin{equation}
U_D = k_{B}T\ln(c_D/c) \label{contact}
\end{equation}
in the channel, and prevent remaining K cations from leaving the
channel. The contact potential appears because negative charges in
the channel are immobile (belong to practically static DNA). In this
sense this contact potential is similar to Donnan potential
appearing on membranes permeable only for one sign of ions. This
contact potential is also similar to the contact potential at the
junction between a $p$-type doped and an intrinsic semiconductor.

The contact potential moves down energies of both bands of extra
cations and vacancies, while bending these bands up in the very ends
(Fig.~\ref{figband}b). On the other hand, the energy band of an
extra anion (shown in Fig.~\ref{figband}b by the dashed line) is
moved up by the contact potential. (Without contact potential this
band coincided with the energy band of an extra cation as shown in
Fig.~\ref{figband}a). This leads to the total exclusion of anions
from the channel. Such exclusion was also noticed in Ref.
~\cite{Rabin}.

At $c > c_D$ both additional cations and anions freely enter the
channel in equal number from the bulk in order to equilibrate
their concentration. This does not lead to the contact potential
because this process does leave behind layers of fixed charges. At
$c > c_D$ ion current through the channel should be due to both
extra salt cations and anions and, therefore, should be
proportional to salt concentration $c$. On the other hand, at $c <
c_D$ the transport is due to cations only and current is roughly
independent on $c$, because $U_D$ grows with decreasing $c$ and
the contact potential $-U_D$ reduces the barrier $U(0)$
(compensating for decreasing $c$). One can say that at small $c$
transport is only due to cations neutralizing DNA. For example,
one of them residing near the right end of the channel may go
through channel to left end, while another cation from the left
solution replaces it.

These ideas provide reasonable interpretation for the experimental
data~\cite{Bonthuis} for the blocked ion $I_{b}(c)$ as a function
of salt concentration $c$. It was found that $I_{b}(c) \propto c$
at $c \geq 1$ M, while at at $c < 1$ M the current $I_{b}(c)$
weakly depends on $c$. We interpret this data as evidence that
$c_D \sim 1$ M. (See more about the experimental data for
$I_{b}(c)$ and their explanation in Ref.~\cite{Bonthuis}.) We are
not trying here to estimate $c_D$ microscopically because this
would require dealing with ion sizes.)

\section{Effective charge of the stall force}
\label{sec_qs}

The effective charge $q_s$ is defined by Eq.~(\ref{fpqp}) for the
force $F_s$ necessary to stall ssDNA, when the ssDNA occupies the
whole channel (see Fig.~\ref{figdna}). We show below that the
stall charge $q_s$ is proportional to the blocked current $I_b$.
Let us concentrate on the case $c < c_D$, when anions (Cl$^{-}$)
are excluded from the channel and all blocked ion current is due
to cations.

Let us assume that external electric field, $E=-V/L$, is applied
in the direction opposite to $x$-axis. It generates a force
$-N_{0}eE$ on $N_{0}$ ssDNA charges in the channel moving DNA
along $x$-axis. The opposite force $N_{0}eE$ acts on the cations.
If the average drift velocity of DNA is $v_D$, and the average
drift velocity of cations inside the channel is $v_{c}$ we can
write two momentum balance equations of steady state viscous
motion for $v_c$ and $v_D$:
\begin{equation}
N_{0}eE = k_{c}v_{c}+ k_{cD}(v_{c} - v_{D}), \label{frictioncat}
\end{equation}
\begin{equation}
-N_{0}eE + F_a = k_{D}v_{D} + k_{cD}(v_{D}- v_{c}).
\label{frictiondna}
\end{equation}
Here $F_a$ is the additional non-electrostatic force applied to
ssDNA along $-x$ (in the direction opposite to DNA motion),
$k_{c}$ and $k_{cD}$ are friction coefficients of $N_{0}$ cations
with the channel walls and with DNA respectively, while $k_{D}$ is
the friction coefficient of DNA with the channel walls.

Although water is not included in these equations explicitly, it
is water viscosity that provides the transfer of momentum between
moving DNA and cations and from them to the walls. If DNA is long
(compared to $N_{0}$), the friction force of free ends with the
water is also included in $k_{D}$. Note that voltage drops on the
membrane, so that electric field acts only on $N_{0}$ phosphates
and $N_{0}$ cations.

Eq.~(\ref{frictioncat}) and (\ref{frictiondna}) give non-trivial
predictions. First, when $F_a = 0$, addition of
Eq.~(\ref{frictioncat}) and (\ref{frictiondna}) gives
\begin{equation}
v_{D}= - k_{c}v_{c}/k_{D}. \label{theorem}
\end{equation}
So the average drift velocity of ssDNA is proportional to average
drift velocity of cations. In experiment both $v_{c}$ and $v_{D}$
are small, so that the linear in $v_{c}$ and $v_{D}$ approximation
is justified. Formally, Eq.~(\ref{theorem}) looks like a drag of
DNA by cations. Actually, this is an anti-drag, because the
coefficient in Eq.~(\ref{theorem}) is negative. The reason for the
anti-drag is simple: if $k_c$ or $v_c$ were equal to zero, the
$N_0$ cations would transfer all momentum they receive from the
electric field to the DNA molecule. Then DNA would not move at
all. The transfer of negative cations momentum to walls makes the
net force applied to DNA positive. Thus, DNA moves in $x$
direction only when cations move in $-x$ direction!

Second, when $F_a = F_s$, by the definition the applied force
stalls the ssDNA setting $v_D=0$. Adding Eq.~(\ref{frictioncat})
and (\ref{frictiondna}) at $v_D=0$ we find the stalling force
\begin{equation}
F_s = k_{c}v_{c}, \label{stallingforce}
\end{equation}
where strictly speaking $v_c$ is the average drift velocity of
cations at $v_D=0$. Actually in all experiments $v_D \ll v_c$,
which means $k_D$ is very large. Thus, one can use $v_c$ from
experiments where $F_a=0$. This allows us to express $v_c$ through
blocked current $I_b$, which may be written as
\begin{equation}
I_b = ne v_c, \label{Ib}
\end{equation}
where $n = N_0/L$ is the linear density of cations in the channel.
Both $I_b$ and $v_c$ are small because of electrostatic barrier
for ion motion in the blocked channel. Indeed, most of the time
all cations are bound to DNA phosphates. Only rarely a cation goes
through middle of the channel contributing both into the current
and the drift velocity $v_c$.

Combining Eq. (\ref{stallingforce}), (\ref{fpqp}) and (\ref{Ib})
we arrive at at
\begin{equation}
q_s = -k_{c}I_{b}/ne E. \label{stallingcharge1}
\end{equation}
We can exclude $E$ from Eq.~(\ref{stallingcharge1}) using equation
for the current of the open channel
\begin{equation}
I_0\!= \!n_{0} e v_c^{0} = \!n_{0}e^2E/6\pi \eta R. \label{I0}
\end{equation}
Here $v_{c}^{0}= eE/6\pi \eta R$ is the drift velocity of a cation
in open channel, $n_{0}=2c\pi a^2 $ is a number of cations and
anions per unit length of the open channel (we assume that they have
same radius $R$ and use Stokes formula for viscous resistance force
because $R\ll a$). Combining Eq. (\ref{stallingcharge1}) and
(\ref{I0}) we get
\begin{equation}
q_s = - \frac{k_{c}e}{6\pi \eta R}\frac{n_{0}I_{b}}{n I_0}.
\label{stallingcharge2}
\end{equation}
The last step is to calculate $k_c$. When the cation moves in the
blocked channel it transfers the force $eE$ half to the wall and
half to the DNA. We write the force on the wall as $6\pi \eta R
v_c\cdot f(R/b)$. If the cation size is small enough $R<<b$ the
Stokes formula is applicable and $f(R/b)=1/2$. If the cation is
large so that $b-2R<<b$ one arrives at the asymptotic estimate
$f(R/b) = \beta \ln {2R \over b-2R}$, where according to
Ref.~\cite{Brenner} the coefficient $\beta = 8/15$. Here we
illustrate the origin of this logarithmic expression by the
following simple derivation leading to slightly different $\beta =
2/3$. Let us locally approximate both the DNA surface and internal
walls of the channel as static parallel plane walls at $z = 0$ and
$z = b$ (see Fig.~\ref{figfield}). Let us assume that center of the
ball is moving in the plane $z = b/2$ with velocity $v_c$. When the
ball is at $ x = y = 0$ and $z$ axis is a polar axis of the ball let
us cut the spherical surface of the ball by a big number of coaxial
cylinders with the axis $z$ and radiuses $\rho$ in the range $0 <
\rho < R $. We arrive at a number of rings on the surface of the
ball. Let us estimate the force to the external wall as a sum over
close to it rings. A distance of the ring with the radius $\rho \ll
R$ to the wall is $ h(\rho) = (b/2) - R + \rho^{2}/2R$, the local
gradient of velocity is $\sim v_{c}/h$, the effective area of the
ring is $2 \pi \rho d\rho$ and the total momentum transfer rate
(force) is
\begin{equation}
2 \pi \eta \!\int_0^R \frac{v_c\rho d \rho }{h(\rho)} \simeq 4\pi
v_c \eta R \!\int_{b/2\!-\!R}^{R}\!\frac{d\rho}{\rho} = 4\pi v_c
\eta R \ln {2R \over b\!-\!2R} \label{momentum}
\end{equation}
This leads to $f(R/b)\simeq \beta \ln {2R \over b-2R}$ with $\beta=
2/3$ for a cation in middle plane $z=b/2$. For a cation closer to
external wall at finite $b/2 < z < b-R $ we get $\ln{2R \over
b-R-z}$ instead of $\ln{2R\over b-2R}$. Because dependence on $z$ is
only logarithmic, averaging of $f$ over the cation positions $z$
makes a negligible change to $\beta$. Assuming that all $N_0$
cations move with average drift velocity $v_c$ we find that the
total force cations exert on the wall is $6\pi N_0 \eta R
v_{c}f(R/b)$. This gives $k_c = 6\pi N_0 \eta R f(R/b)$ and
\begin{equation}
q_s = -eN_{0}\frac{n_{0}I_{b}}{n I_0}f(R/b).
\label{stallingcharge}
\end{equation}
For ssDNA in $\alpha$-HL channel using $\beta = 8/15$ and a crude
estimate $\ln{2R \over b-2R} \sim 1.8 $ we arrive at $f(R/b)\simeq
1$. The ratio $n_0/n \sim 1$ at the typical salt concentration $c =
1$~M. Thus, we arrive at Eq.~(\ref{main}) and $q_s \simeq -1e$.

This may look surprising because the net DNA charges of the DNA
contact layers near the two channel ends together are definitely
larger than $e$. These charges, however, do not move with DNA and do
not contribute to the pulling DNA force and to the effective charge
$q_s$ of the stall force.

Up to now we dealt with the case of relatively small
concentrations of salt, $c < c_D$, when linear concentration of
cations $n$ inside the blocked channel is fixed and anions are
excluded from the channel. One can show that at $c > c_D$ when
current is due to both cations and anions in
Eq.~(\ref{stallingcharge}), one should drop the ratio $n_{0}/n$.

We are not aware of any direct measurement of $q_s$, for ssDNA in
$\alpha$-HL channel. In the next two sections we show how one can
express the capture charge $q_c$, the unzipping charge $q_u$ and
the escape charge $q_e$ through the stall charge $q_s$.

\section{Effective charge of the capture rate}
\label{sec_qc}

\begin{figure}[ht]
\begin{center}
\vspace{-2.5cm}
\includegraphics[height=0.5\textheight]{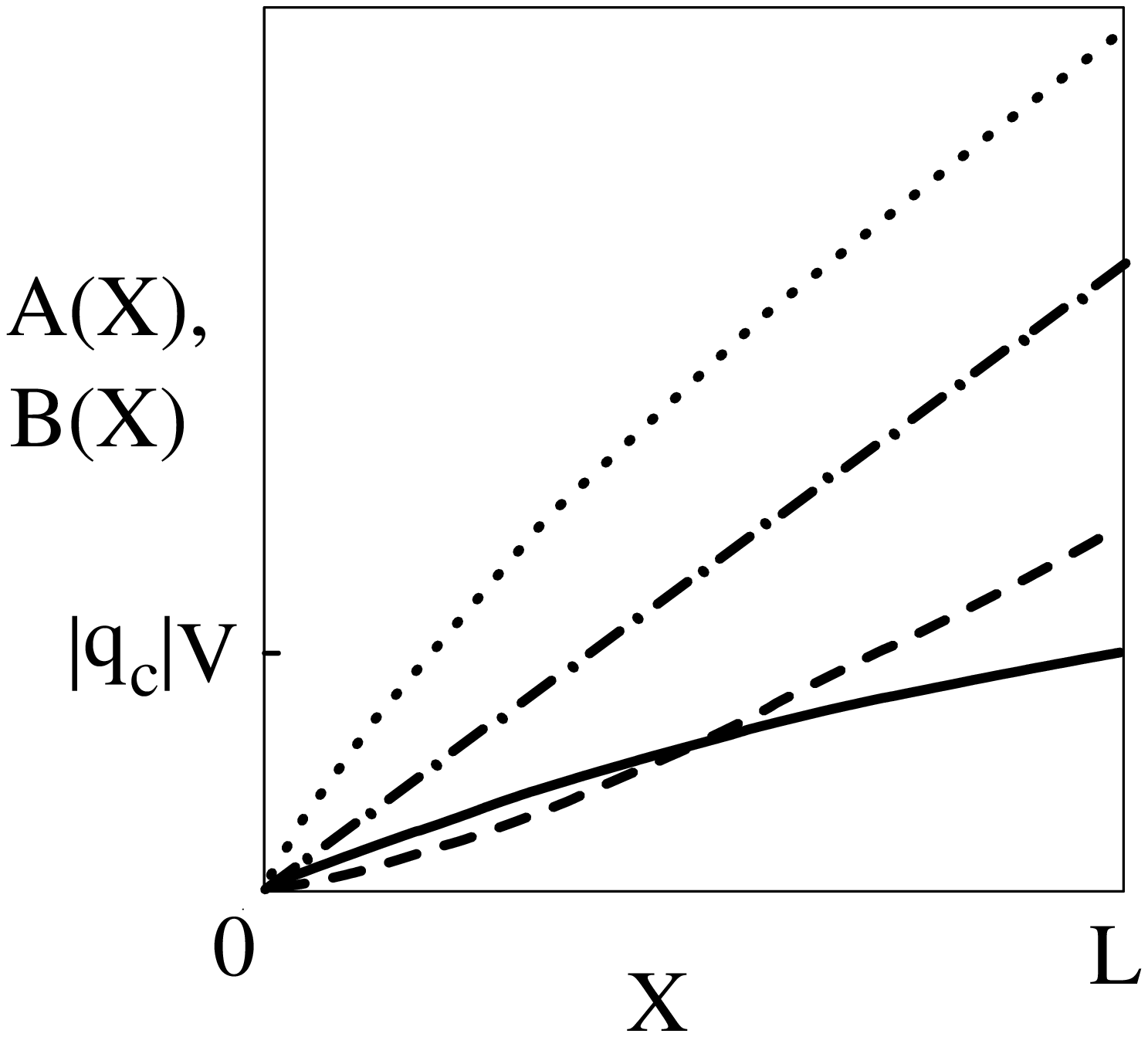}
\end{center}
\vspace{-2.5cm} \caption{Schematic plots of several free energy
barriers $B(X)$ for the partial (to the length $X$) DNA capture.
The barrier for DNA capture without electric field is nearly
linear (dash-dotted line). Also shown are the barriers $B(X)$ for
the direction in which $V$ helps to capture DNA (dashed line) and
for the direction in which $V$ hinders the capture (dotted line).
The work $A(X)$ done by the applied potential $V$ is shown by
solid line.} \label{figcapbarrier}
\end{figure}

In the preceding section we focused on the stall force $F_s$ or
pulling force $F_{p} = - F_{s}$ in the situation, when the ssDNA
already occupies the whole channel (Fig.~\ref{figdna}). In order
to calculate the effect of the applied voltage on the capture
rate, we turn to the pulling force $F_{p}(X)$ in the situation,
when the ssDNA penetrates only a fraction of the channel $X/L$
(Fig.~\ref{figdnapart}). The length of DNA in the channel $X$
changes from $0$ to $L$ while DNA enters the channel, and the work
done by the force pulling ssDNA is
\begin{equation}
- q_{c}V=\int_{0}^L F_p(X)dX, \label{work}
\end{equation}
This work reduces the free energy barrier of the capture rate
(see Fig.~\ref{figcapbarrier}). Below we assume that all voltage
drops on the blocked part of the channel, because unblocked part
of the channel is so wide that its resistance is much smaller than
that of the blocked part. Then, similarly to
Eq.~(\ref{stallingcharge1}), the force $F_p(X)$ can be expressed
through the blocked current of the partially blocked state
$I_b(X)$,
\begin{equation}
F_p(X)= - { k_cI_b(X)\over ne}. \label{proportion2}
\end{equation}
The force $F_p(X)$ is larger than $F_p(L)$, because a shorter
blocked channel leads to a larger current $|I_b(X)|
 >| I_b(L)|$ at a given voltage $V$. The main reason is the smaller
electrostatic barrier for the traversing cations.

Combining Eq.~(\ref{work}), (\ref{proportion2}) and
(\ref{stallingcharge1}) we arrive at a simple result:
\begin{equation}
{q_{c}\over q_{s}}={1\over L}\int_{0}^L {I_b(X)\over I_b(L)}dX.
\label{ratio}
\end{equation}
In order to calculate this ratio for ssDNA in $\alpha$-HL channel
we use the experimental data~\cite{Meller2001} for the blocked
current $I_{b}(N)$ as a function of the total DNA length (in
bases), when it is shorter than $N_0 = 12$, namely $4 \leq N \leq
12$. We assume that $I_{b}(N)$ can be used for $I_{b}(X)$ in
Eq.~(\ref{ratio}). In this way we obtain $q_{c}/q_{s} \approx 2.5
\pm 1.0$ somewhat larger than the experimental value $q_{c}/q_s
\approx 1.9e/ (1.1e)=1.7$. We emphasize that in agreement with our
arguments both theoretical and experimental values are
substantially larger than $e$.

In Fig.~\ref{figcapbarrier} we illustrate the difference between
capture barriers without a voltage, along the voltage pulling
force and against this force.

\section{Voltage enhanced hairpin unzipping and escape against the voltage.}
\label{sec_qu}

In this section we deviate from the capture rate theory and
discuss relationship between unzipping and escape effective
charges $q_u$ and $q_e$ defined in Introduction and the
fundamental stall charge $q_s$. The charge $q_u$ determines the
voltage dependence of the release rate of ssDNA, when it is
trapped in the channel by an intentionally designed double helix
DNA hairpin at the end (see Eq.~(\ref{escaperate})). Double helix
DNA of the hairpin is too thick to go through the $\alpha$-HL
channel. Thus, unzipping of the hairpin is necessary in order to
release DNA from the channel to the right (Fig.~\ref{fighairpin}).

In experiment~\cite{Mathe} ssDNA was first inserted into the
$\alpha$-HL channel and then kept in by a relatively low voltage.
The voltage was increased to the large value $V$ at time $t_0$ and
probability that DNA is still in the channel at the time $t_0 + t$
was measured. This probability behaves as $\exp (- R_{u}t)$
defining the rate of unzipping $R_{u}$ in Eq.~(\ref{escaperate}).

The rate $R_{u}$ was found~\cite{Mathe} to grow with the voltage
according to Eq.~(\ref{escaperate}). This equation defines
unzipping charge $q_{u}$. We would like to show that $q_{u}=
q_{s}(M/N_0)$, where $M$ is the number of base pairs in the
hairpin. (In the experiment~\cite{Mathe} $M = 7, 9, 10$).


%
\begin{figure}[ht]
\begin{center}
\vspace{-4.5cm}
\includegraphics[height=0.6\textheight]{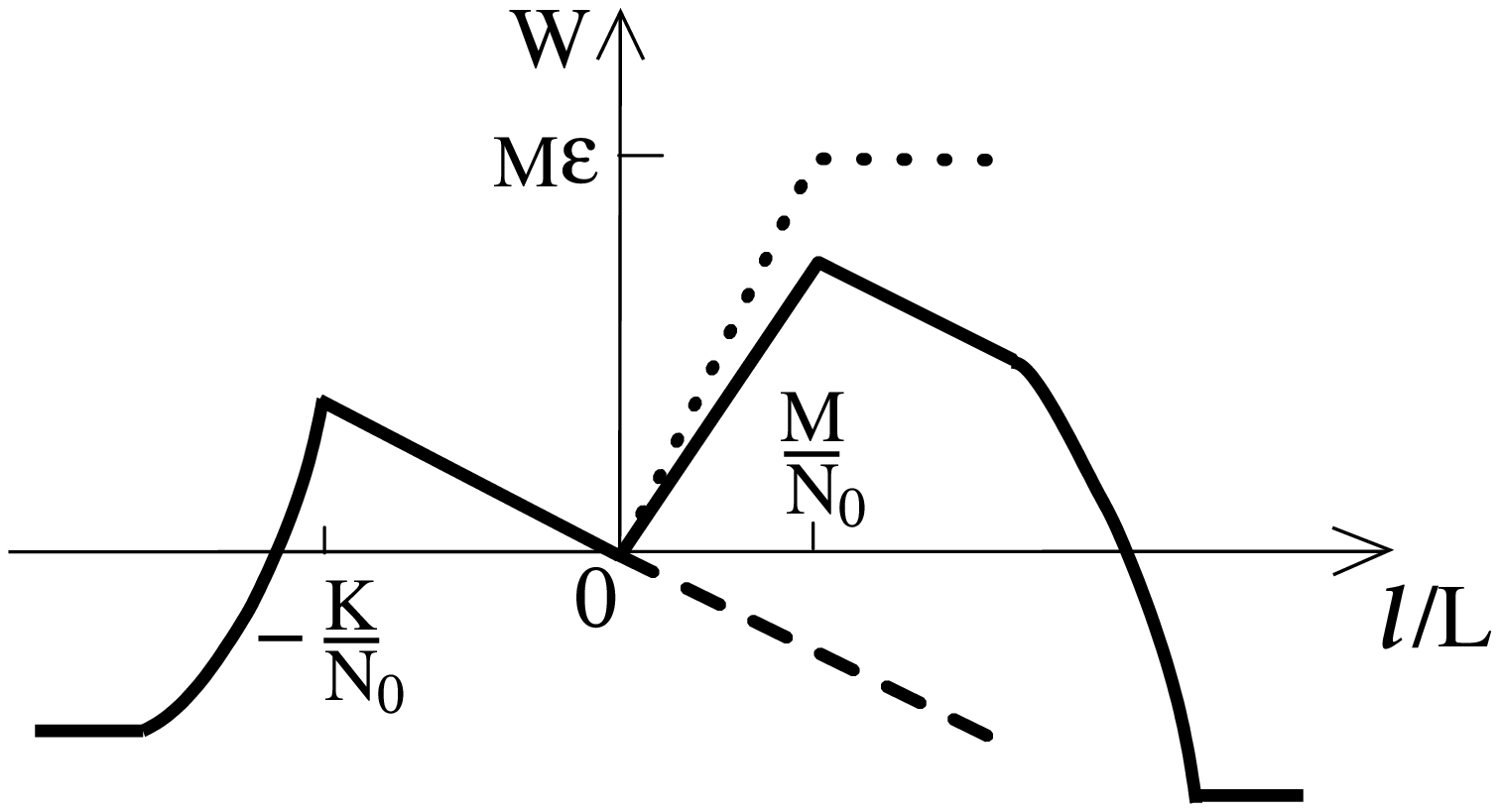}
\end{center}
\vspace{-4.5cm} \caption{Free energy $W$ of ssDNA with a hairpin
as a function of displacement $l$ of DNA (solid line). Positive
$l$ correspond to unzipping and release to the right, and for this
case $l$ is defined as the length of DNA passing by the right end
of the channel starting from original configuration of
Fig.~\ref{fighairpin}. Negative $l$ corresponds to the DNA escape
in the direction of hairpin, for this case $-l$ is defined as the
length of DNA passing by the left end of the channel
(Fig.~\ref{fighairpin}). The dotted line is the energy of broken
base pairs, the dashed line is the energy $W = -F_{p}l =
q_{s}Vl/L$ of the voltage induced pulling force. Relative height
of the barriers for unzipping and escape apparently depends on the
magnitude of the voltage $V$.} \label{figpotential}
\end{figure}
%

Indeed, unzipping would raise DNA eventually to the barrier $M
\varepsilon$, where $\varepsilon$ is energy of one base pair. This
happens when DNA moves along $x$-axis by $M$ bases or by the
length $l_{max} = ML/N_0$. Corresponding potential energy is shown
in Fig.~\ref{figpotential} by dotted line as a function of
displacement $l > 0 $ of the DNA. The electric field reduces the
barrier. After unzipping and moving $2M$ hairpin bases through the
channel, DNA slides down the capture barrier (see
Fig.~\ref{figcapbarrier}) and eventually reaches a constant energy
plateau in the bulk of the right solution
(Fig.~\ref{figpotential}).

In order to evaluate the correction to the barrier we should
assume that DNA moves very slowly and calculate the work of the
pulling force $F_p = - q_{s}(V/L)$, along the displacement
$l_{max} = ML/N_0$. This gives the correction to the barrier
$F_{s} l_{max} = |q_{s}|(M/N) V = |q_{u}|V$, and, therefore
\begin{equation}
q_{u}= q_{s}\frac{M}{N_0}. \label{unzipping}
\end{equation}
We estimated above that $ q_{s} \sim -1e$. In experiments
~\cite{Mathe} $M/N_0 \sim 1$ and we get $q_u \simeq -1e$. This is
close to $q_{u} \simeq -1.1e$ found in Ref. ~\cite{Mathe}.

Our derivation is valid when the unzipping barrier corrected by
voltage $V$ is still much larger than $k_BT$. In this case even
with applied force, unzipping randomly alternates with zipping,
while the saddle point corresponding to the totally unzipped
hairpin is being reached in equilibrium way. To our mind, this
condition is satisfied in the original experiment~\cite{Mathe} and
therefore, agreement of our theory with its results can be
expected \cite{foot2}. More complicated problem of unzipping of
several hairpins of RNA was discussed recently~\cite{Bund}.

Let us switch to the escape effective charge. Suppose DNA with a
hairpin is brought to the channel by a pulling voltage
(Fig.~\ref{fighairpin}). Let us assume that single stranded part
of DNA has $N_0 + K$ bases, so that $K$ bases have already arrived
to the bulk solution opposite to the hairpin. Let us keep DNA in
the channel by an applied voltage $V$, which is so small that the
unzipping rate is smaller than the rate of escape in the direction
opposite to the pulling force. The rate of such alternative escape
should behave according to Eq.~(\ref{relrate}) and dominate at
small enough voltages. We want to show that the escape charge
$q_{e}= |q_{s}|(K/N_0)$. The free energy profile $W$ for the
escape process (negative $l$) is shown in Fig.~\ref{figpotential}
together with the unzipping process (positive $l$). The pulling
force preventing the escape is $F_p = |q_{s}|V/L$. DNA is climbing
up against this force until all the DNA fits in the channel (no
tail in the bulk solution opposite to hairpin). Starting from this
point DNA slides down the capture barrier before reaching the
energy plateau, when all DNA arrives to the left solution (see
Fig.~\ref{figcapbarrier}). Therefore, displacement at which $F_p$
works is $l_{max}^{*} = - LK/N_0$ and the escape barrier is
$|l_{max}^{*}| F_p = |q_{s}|(K/N_0)V$ or
\begin{equation}
q_{e}= |q_{s}|\frac{K}{N_0}. \label{es }
\end{equation}

The only measurement of $q_e$ we know was done in
Ref.~\cite{Nakane}. It resulted in roughly speaking four times
larger value of $|q_s|$ than we estimated here, but this result
should be taken with caution, because surprisingly in this
experiment $I_{b}$ has unconventional sign. We hope that
measurements of $q_e$ will be repeated.

\section{Electrostatic free energy barrier for DNA capture at low voltages}
\label{sec_barrier}

In this section we return to the capture rate at a small voltage
$V$ and concentrate on the difference between free energies of
screening atmospheres of a rod-like DNA molecule in the channel
and in the bulk solution, contributing into the capture rate
barrier. We deal with the neutral channel, for which self-energy
plays no role. Therefore, in the first approximation one can
neglect discreteness of charges, assuming the DNA surface charges
are uniformly smeared and the distribution of ions obeys
Poisson-Boltzmann equation. To simplify calculation we look on
unfolded DNA surface as an uniformly charged plane at $z=0$ and
the unfolded inner channel wall as another neutral plane at $z=b$
(Fig.~\ref{figfield}). We calculate the free energy price to push
the neutral plane from infinity to the distance $b$ compressing
the screening cloud. It is easy to verify that the price grows as
the distance between the planes $b$ decreases. The fact that the
DNA surface is cylindrical is not important when the wall
separation $b$ is smaller than the DNA radius $r$. We label the
charge density of DNA surface as $-\sigma$, the mean field
electric potential as $\phi$.

To find the free energy of the system we need to study $\phi(z)$
for a given wall separation $b$. Far from the channel ends $\phi$
depends only on $z$, and the Poisson-Boltzmann equation for such a
system for $0<z<b$ is
\begin{equation}
{d^2 \phi \over dz^2} = {8 \pi ec\over \kappa}
\sinh\left({e\phi\over k_B T}\right) , \label{possionboltzmann}
\end{equation}
It should be solved with the boundary conditions which follows
from the over-all neutrality and the absence of charges at $z<0$
and $z>b$
\begin{equation}
{d \phi \over dz}|_{z=b} = 0 ,\,\,\,\,\,\,\,\,\,\,\,\, {d \phi
\over dz}|_{z=0} = {4 \pi \sigma \over \kappa}. \label{atwall}
\end{equation}
\begin{figure}[ht]
\begin{center}
\includegraphics[height=0.22\textheight]{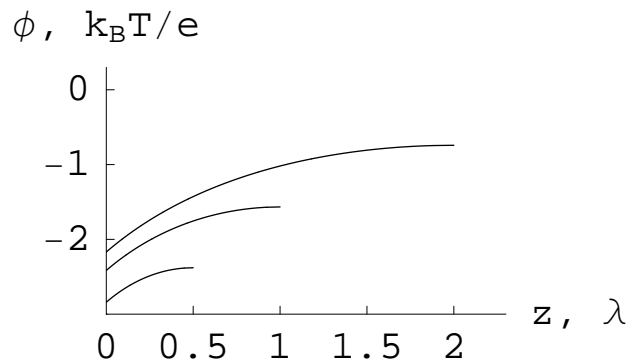}
\end{center}
\caption{ The potential $\phi$ as a function of $z$ for
$b/\lambda=0.5$, $1$ and $2$ (from bottom to top), at ion
concentration $ce^2/(l_B\sigma^2)=1$. Here the Gouy-Chapman length
$\lambda= k_B T \kappa / (2\pi \sigma e)$ is the unit of distance. }
\label{figfih}
\end{figure}

The integration of Eq.~(\ref{possionboltzmann}) gives
\begin{equation}
{d\phi\over dz} = \sqrt{{16\pi k_B Tc\over \kappa}
\cosh\left({e\phi\over k_B T}\right)+A}. \label{integration1}
\end{equation}
With the help of the second boundary condition (\ref{atwall}) we
can write $A$ in the form $A=({4 \pi \sigma \over
\kappa})^2-{1\!6\pi k_B Tc\over \kappa} \cosh\left({e\phi(0)\over
k_B T}\right)$, and calculate numerically
\begin{equation}
z(\!\,\phi\!\,)\!=\!\!\int\limits_{\phi(0)}^{\phi} \!\! {d\psi
\over \sqrt{{1\!6\pi k_B T c \over \kappa} [\cosh({e\psi\over k_B
T})\!-\!\cosh({e\phi(\!\,0\!\,)\over k_B T})]\!\!+\!\!(\!{4 \pi
\sigma \over \kappa}\!)^{^2}} }, \label{integration2}
\end{equation}
Numerical inversion of Eq.~(\ref{integration2}) gives $\phi(z)$.
Then the only remaining parameter, the constant $\phi(0)$ in the
integral Eq.~(\ref{integration2}) is determined using the first
boundary condition (\ref{atwall}).
\begin{figure}[ht]
\begin{center}
\includegraphics[height=0.22\textheight]{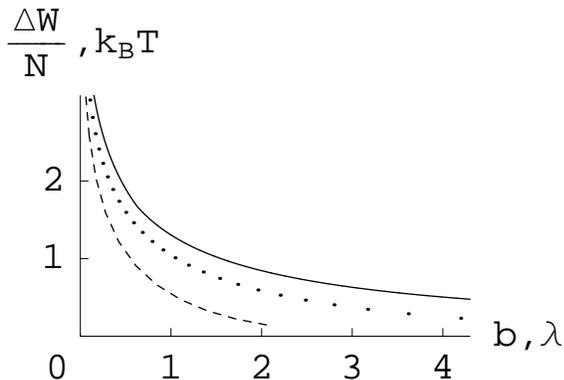}
\end{center}
\caption{Free energy of the capture barrier per one screened DNA
charge in the channel as a function of $b$ (in units of $\lambda$)
for $c=0$ (solid line), $ce^2/(l_B\sigma^2)=0.1$ (dotted line)and
$ce^2/(l_B\sigma^2)=1$ (dashed line).} \label{figFh}
\end{figure}
In this way we can find the potential $\phi(z)$ for any given wall
separation $b$. A few examples are shown in Fig.~\ref{figfih}.

We emphasize that the results obtained so far are valid for the
neutralized main part of the ssDNA in the channel. They are not
applicable to the DNA charges near the end of the channel, which
lose their counter ions to the bulk solution forming two contact
layers. It is the contact potential $-U_D$ that makes the
potential $\phi(z)$ in the neutral part of the channel negative
even at $z=b$ (see Fig.~\ref{figfih}). This is not surprising,
because through Eq.~(\ref{integration1}) the potential $\phi(z)$
is related to $c$, which is in turn directly related to $U_D$ by
Eq.~(\ref{contact}).

The total free energy of the system per unit area of DNA surface
can be calculated as
\begin{equation}
{W\over area}\!=\!{\!-\sigma \over 2}\phi(0)+\!\sum_{\pm}
\int\limits_{0}^{b} \!\left[{\pm e\over 2}\phi(z)\!-\!k_B
T\ln{c\over c_{\pm}}\right]\!c_{\pm}dz. \label{freeenergy1}
\end{equation}
The Boltzmann distribution
\begin{equation}
c_{\pm}=c \exp[-{{\pm}e\phi(z)\over k_B T}] \label{freeenergy}
\end{equation}
reduces Eq.~(\ref{freeenergy1}) to
\begin{equation}
{W\over area}=c \int\limits_{0}^{b} \sinh\left[{e\phi(z)\over k_B
T}\right] e\phi(z)dz-{\sigma \over 2}\phi(0). \label{freeenergy2}
\end{equation}
One can find the free energy at any given $b$ and plot it choosing
$b=\infty$ as the reference point (Fig.~\ref{figFh}).

For ssDNA threading through $\alpha$-HL channel with
$\sigma=Ne/[\pi (r+a) L]$, the Gouy-Chapman length $\lambda= {k_B
T \kappa \over 2\pi \sigma e}$, the salt concentration $c$ in
units of
$l_B\sigma^2/e^2$, and free energy per area in units of
$k_B T\sigma/e$. Here $l_B={e^2\over \kappa k_B T}$ is the Bjerrum
length (for water at the room temperature $l_B=0.7\,$nm).
\begin{figure}[ht]
\begin{center}
\includegraphics[height=0.17\textheight]{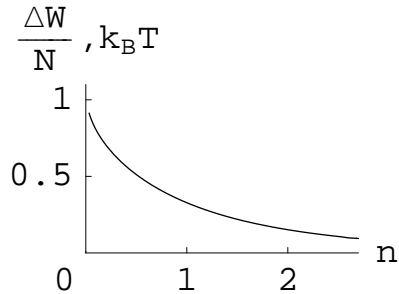}
\end{center}
\caption{Free energy barrier per one of screened DNA charge in the
channel as a function of the dimensionless salt concentration $n =
ce^2/(l_B\sigma^2)$, for the wall separation $b=1.4\,\lambda$.}
\label{figFn}
\end{figure}

For the system of ssDNA threading through $\alpha$-HL channel, the
Gouy-Chapman length $\lambda$ is about $0.25\,$nm, while the wall
separation is about $0.35\,$nm. The free energy barrier as a
function of salt concentration in this example is shown by the
solid line in Fig.~\ref{figFn}.

We see that for the salt concentration $c=1\,$M or
$n=ce^2/(l_B\sigma^2)=1.37$, the barrier is about $\Delta W=0.24k_B
T \cdot N\simeq 2.4\,k_B T$. Here $N = N_0 - 2N_D$ is number
screened DNA phosphates in the channel and $N_D$ is number of
phosphates in each contact layer. We estimated that $N_D\sim 1$ and
therefore, $N = 12 - 2 =10$ in the range of salt concentrations
between $0.25\,$M and $c=1\,$M. For smaller salt concentration
$c=0.5\,$M and $c=0.25\,$M, the barriers are $4.3\,k_B T$ and
$6.0\,k_B T$. The smaller $c$ the larger barrier. This barrier being
added to conformation entropy barrier of ssDNA discussed in
Introduction may invert dependence of the total barrier on the salt
concentration and explain the observed growth of the capture rate
with the salt concentration.

\section{Conclusion}
\label{sec_conclusion}

In this paper we evaluate effective charges of DNA responsible for
the pulling and stall forces, for the voltage affected capture
rate, as well as for voltage induced unzipping ssDNA with a
hairpin and for its escape against the pulling force of the
electric field. The stall charge $q_s$ is the most fundamental
one, measurements of the other three charges can be used to
evaluate it.

The main result of this paper is the linear equation connecting
the stall charge with the blocked ion current $I_b$. In the
simplest form of Eq.~(\ref{stallingcharge}) it is applicable only
for relatively small concentration of salt, $c < c_D$, when
blocked current is due to neutralizing cations only and,
therefore, roughly speaking is salt concentration independent.
This equation is based only on momentum conservation, does not
depend on the mechanism of the ion current blockage or specific
model of DNA and, therefore, has a high degree of universality.

We also find a new kind of the barrier for DNA capture, which can
explain the puzzling growth of the low voltage capture rate
$R_c(0)$ with the salt concentration $c$. We show that such a
barrier results from squeezing of the screening cloud of DNA when
DNA enters the channel.

The focus of this paper is on ion channels or nanopores barely
permitting DNA translocation. Our theory can be applied to ssDNA
translocating through a solid state nanopore with diameter
comparable to $\alpha$-hemolysin~\cite{Min}. It should also work
for a double helix DNA in a solid state nanopore, which \ diameter
$2a \leq 3$~nm only slightly exceeds the diameter of DNA (2~nm).
For a wider pore with $2a \geq 4$~nm the self-energy electrostatic
barrier vanishes and one can use purely hydrodynamic theory of the
stall charge~\cite{Long,Lemay}.

\begin{acknowledgments}
We are grateful to R. Bundschuh, A. Yu. Grosberg, A. Kamenev, S.
G. Lemay, A. Meller, Y. Rabin, A. M. Tsvelik for useful
discussions. B. S. is also grateful for hospitality to KITP and
ACP, where this paper was finished.
\end{acknowledgments}



\begin{thebibliography}{100}
\bibitem{foot} Actually, at finite ratio $\kappa/\kappa'=40$ some electric
lines leak from the channel. In empty cylindrical channel of
radius $a$ this happens with characteristic length $\xi = 6.8a$.
ss-DNA occupies roughly speaking $40\%$ of the channel
cross-section area. This leads to $20\%$ smaller effective radius
of the water filled channel and correspondingly to somewhat
stronger leakage and smaller $\xi \sim 5.5a$. Leakage of electric
field reduces the electrostatic barrier~\cite{Zhang} by the factor
$(2\xi/L)\tanh(L/2\xi)\simeq 0.88$. Below we neglect difference
between this factor and unity and ignore all the leakage effects.
\bibitem{foot1} The entropy effect is, however, much weaker than
in the one-dimensional case studied in Refs.~\cite{Zhang,Kamenev}.
As we see in Sec.~\ref{sec_Electrostatics} in the $\alpha$-HL
channel blocked by ssDNA the self-energy has a local logarithmic
contribution $U_{2}(0)$ of the two-dimensional origin. It leads to
additional small factors in the important dimensionless constants
$\alpha$ and $\gamma$ in Refs.~\cite{Zhang,Kamenev}, reduces them to
small enough values, and as a result the electrostatic barrier
survives. For wider nanopores with diameter $a \sim 4$~nm blocked by
a double helix DNA~\cite{Min} with the diameter $2$~nm, the width of
the water-filled space between DNA and channel wall $b = 1$~nm is
larger than $l_B$. In this case, $U_{2}(0)$ is not important and at
$c =1$~M the electrostatic barrier should vanish~\cite{Zhang,
Kamenev}. Indeed, the blocked ion current $I_{b} = 0.63I_{0} $ what
is not too far from the fraction of the blocked cross-section.
\bibitem{foot2} In the later paper~\cite{Mathe06} the authors
treated experimental data in a different model and have got twice
larger charge $q_{u} \simeq -2.2e$. We do not understand the
origin of this difference.
\end{thebibliography}
\end{document}